\documentclass[12pt,twoside]{article}
\usepackage{fleqn,espcrc1}

\usepackage{graphicx}
\usepackage[figuresright]{rotating}

\newcommand{\NP}[1]{ Nucl.\ Phys.\ {#1}}

\newcommand{\PL}[1]{ Phys.\ Lett.\ { #1}}

\newcommand{\PR}[1]{Phys.\ Rev.\ { #1}}
\newcommand{\PRL}[1]{ Phys.\ Rev.\ Lett.\ { #1}}

\newcommand{\AmS}{{\protect\the\textfont2
  A\kern-.1667em\lower.5ex\hbox{M}\kern-.125em}}
\newcommand{\vba}{\not{\!v}}
\newcommand{\partba}{\not{\!\!\partial}}
\title{The Inverse Amplitude Method and Heavy Baryon Chiral Perturbation
Theory applied to pion-nucleon scattering
\thanks{Talk given at the 8th International Conference on 
Hadron Spectroscopy, HADRON99,
August 24-28, 1999, Beijing, China.
Work partially supported by DGICYT under contract
AEN97-1693.}}

\author{ J. R. Pel\'aez$^a$ and A. G\'omez Nicola
  \address{ Departamento de F\'{\i}sica Te\'orica.
Universidad Complutense. 28040 Madrid. SPAIN.},
}
       
\begin{document}

\maketitle

\begin{abstract}
We report on our present work, where by means of the Inverse Amplitude Method
we unitarize the elastic pion nucleon  scattering amplitudes
of Heavy Barion Chiral Perturbation Theory
at $O(q^3)$. We reproduce
the scattering up to the inelastic thresholds including
the $\Delta(1232)$ resonance. The fitted chiral constants
are rather different from those obtained by fitting the extrapolated
threshold parameters for the non-unitarized theory.
\end{abstract}

\section{Introduction: Heavy Baryon Chiral Perturbation Theory}

Chiral Symmetry is a relevant constraint in the interactions between
pions and nucleons, as it is well known since current algebra appeared
in the sixties. However, in order to go beyond the current algebra results
or tree level  calculations from simple models, one needs an
effective theory with a systematic
power counting. Heavy Barion Chiral Perturbation Theory (HBChPT) 
\cite{HBChPT} is
an Effective Theory of nucleons and mesons, built
as an expansion in small momentum transfer (and meson masses) compared
with the typical mass of the nucleons and the chiral symmetry breaking
scale of pions interacting in loops, 
$\Lambda_\chi=4\pi f_\pi\simeq 1.2$ GeV, where $f_\pi$
is the pion decay constant.

When dealing with baryons in an effective Lagrangian
context, the main  difficulty is that  
 the nucleon four momentum is of the same order as the expansion scale, 
since its mass is $m_B\simeq\,1\,\hbox{GeV}$ no matter how small is 
the momentum transfer, and even in the chiral limit \cite{Gasser}.
HBChPT overcomes this problem by treating the baryon fields
as static heavy fermions consistently with Chiral Symmetry,
following the 
ideas of Heavy Quark Effective Theory \cite{Georgi}.
An slightly off-shell baryon momentum can be written as:
\begin{equation}
  p^\mu=m_B\,v^\mu+k^\mu, \quad \hbox{with}\quad v\cdot k\ll m_B.
\label{velocity}
\end{equation}
Then, the Lagrangian ${\cal L}_v$ is given in terms of velocity dependent
baryon fields
\begin{equation}
B_v(x)= \frac{1+\vba}{2}
\exp\left(im_B \vba  v_\mu x^\mu \right) B(x)
\label{Bdef}
\end{equation}
satisfying now a {\em massless} Dirac equation $\partba\, B_v=0$
and whose momenta is $k\ll m_B$. Lorentz invariance is ensured
by integrating 
with a Lorentz invariant measure, i.e. 
${\cal L}=\int\frac{dv^3}{2v^0}\,{\cal L}_v$.
Once this is done, it is possible to find a {\em systematic} power counting 
in $k/\Lambda_\chi$, $k/m_B$, $M/\Lambda_\chi$ and $M/m_B$, where $M$ is the 
mass of the mesons. Generically we will denote $M$ and $k$ by $q$.

In the minimal formulation, only the fields of the
pseudoscalar meson octet and the baryon octet are used to build the effective
Lagrangian. In other cases, the baryon decuplet is also considered as a fundamental field
of the effective Lagrangian. 

With the vertices of the Effective Lagrangian of a given order,
it is possible to calculate Feynman diagrams
containing loops. Each loop increases the order of the diagram so that 
any divergence can be absorbed in the coefficients of higher order operators.
It is therefore possible to
obtain finite results order by order in HBChPT, but
paying the price of more and more chiral parameters.

\section{$\pi$-N scattering in HBChPT}

Despite its difficulty,
there are some one-loop  $O(q^3)$ HBChPT calculations in the
literature  \cite{Mojzis,Meissner}. For $\pi$-N scattering only
four $O(q^2)$ and five $O(q^3)$ chiral parameters are relevant.
In table I we list their estimated values
 obtained from a fit to the
nuclear $\sigma$-term, the Goldberger-Treiman discrepancy 
and ten extrapolated threshold parameters \cite{Mojzis}. 
With this knowledge it was
possible to make six predictions of threshold parameters,
which is ``Neither too impressive nor discouraging'' \cite{Mojzis}. Let us remark
the  rather slow convergence of HBChPT, ``since
the contributions of the first three orders are frequently comparable''.
Encouraged by these results and by the success of unitarization techniques 
in meson-meson scattering \cite{IAM,oop},
our aim is to extend the applicability of HBChPT to 
higher energies implementing unitarity.

\begin{table}[t]
\caption{Values of the chiral constants from a fit of HBChPT to {\em extrapolated}
threshold parameters \cite{Mojzis} and from a IAM fit to phase shifts.}
\begin{tabular}{|c|c|c|c|c|c|c|c|c|c|}
\hline
&$a_1$ & $a_2$ & $a_3$ & $a_5$ & $b_1+b_2$ & $b_3$& $b_6$ 
& $b_{16}-b_{15}$ &$b_{19}$ \\ \hline
HBChPT&-2.6 &1.4 &-1&3.3&2.4&-2.8&1.4&6.1&-2.4 \\ \hline
IAM &9.1 &-8.7 &0.75&27.76&7.07&-9.6&0.056&68.9&-23.13\\\hline
\end{tabular}
\end{table}

As we have seen, within the HBChPT formalism, 
and counting $q_{cm}$ and $M$ as $O(\epsilon)$,
the $\pi$-N amplitudes are obtained
as a series in the momentum transfer.
Customarily  $\pi$-N scattering is described in terms of
partial wave amplitudes of definite
isospin $I$ and angular momentum $J$, that are therefore obtained as
 $t\simeq t_1+t_2+t_3+ O(\epsilon^4)$,
where the subscript stands for the power of $\epsilon$ that 
each contribution carries.  However, an expansion will never satisfy 
{\em exactly} the  $\pi$-N {\em elastic unitarity} condition 
\begin{equation}
  \hbox{Im}\, t = q_{cm}\, \vert t\vert^2,
\end{equation}
although HBChPT satisfies
unitarity {\em perturbatively}. Indeed, we have checked that 
\begin{equation}
  \hbox{Im}\, t_3 = q_{cm}\,\vert  t_1\vert^2.
\end{equation}
Unfortunately, unitarity is a very relevant feature of strong interactions
and it is fundamental in order to incorporate resonances and their associated
poles, which cannot be accommodated in an energy expansion. Note that
although there are other $O(q^3)$ calculations with an explicit $\Delta(1232)$ 
which reproduce phase
shifts up to $E_{cm}\simeq100\,\hbox{MeV}$ \cite{others}, our
aim is the unitarization without including explicit resonance fields.

\section{The Inverse Amplitude Method and $\pi$-N scattering.}

Dividing Eq.(3) by $ \vert t\vert^2$, 
the elastic unitarity condition can be recast as
\begin{equation}
\hbox{Im} (1/t) = -q_{cm} \quad \Longrightarrow \quad 
t\simeq \frac{1}{\hbox{Re}(1/t)-i\, q_{cm}}.
\end{equation}
Any amplitude in this form satisfies
elastic unitarity {\em exactly}. Depending on
how well we approximate the real part of the Inverse Amplitude,
we have different unitarization methods.
For instance, setting $\hbox{Re}(1/t)= \vert q_{cm}\vert\, \hbox{cot}\delta
=-\frac{1}{a}+\frac{r_0}{2}\, q_{cm}^2$ we have the effective range approximation.
If we simply take $\hbox{Re}\, t\simeq t_1$ we obtain a 
Lippmann-Schwinger type equation. Finally, if we
use the  $O(q^3)$ HBChPT expansion we arrive at:
\begin{equation}
t\simeq \frac{t_1^2}{t_1-t_2+t_2^2/t_1-\hbox{Re}\,t_3-iq_{cm}t_1^2},
\label{IAM3}
\end{equation}
which is the $O(q^3)$ form of the Inverse Amplitude Method (IAM). Note that if
we expand again in terms of $q$, we recover at low energies the HBChPT result.

Unitarization methods are not 
foreign to effective theories.
Incidentally, eq.(\ref{IAM3}) is 
nothing but a Pad\'e approximant of the $O(q^3)$ series,
and it is well known that Pad\'es, 
together with very simple phenomenological models are
enough to describe the main features of $\pi$-N scattering.
Although a systematic application with an effective Lagrangian
was demanded, it was never carried out (see \cite{70s} and references therein).
In recent years, with the advent of Chiral Perturbation Theory (ChPT),
the IAM has been applied to meson-meson scattering
with a remarkable success \cite{IAM}. In particular,and
using the $O(p^4)$ ChPT Lagrangians, the IAM reproduces
all the channels up to about 1.2 GeV, including the $\sigma$, $f_0$, $a_0$,
$\rho$, $\kappa$, $K^*$  and octet $\phi$ resonances \cite{oop}.
Very recently, the Lippmann-Schwinger type equation mentioned
above has been applied to S-wave kaon-nucleon scattering with 
eight coupled channels using the lowest order Lagrangian plus one parameter
\cite{or}, reproducing all the low energy cross-sections as well as
the $\Lambda(1405)$ resonance.

\section{Results and Summary}

In Fig.1 we show the preliminary results of an IAM fit to low energy
$\pi$-N scattering phase shifts. It can be noticed that there is a general 
improvement and we obtain a fairly good description up to 
at least the first inelastic threshold for channels even up to
320 MeV of CM energy. Note that the $\Delta(1232)$ resonance 
in the $P_{33}$ channel, has been generated dynamically. Indeed we have
found its associated pole in the
$2^{nd}$ Riemann sheet,  at $\sqrt{s}=1209-i\,46\,\hbox{MeV}$.

The chiral parameters resulting from the IAM fit are given in 
table I. Note how different they are from those obtained in \cite{Mojzis}.
That can be due to several reasons: a) The slow convergence of the series.
Indeed we have checked that contributions from different orders are
comparable in almost every partial wave at the energies we use. 
The effect of higher order terms, which was less relevant at threshold,
is absorbed in our case in the fit of the parameters.
b) Even without unitarization, the values of the parameters 
have a strong dependence on the observables or the
range of energies used to extract them \cite{FeMe}.
c) It could also
suggest that the  $\Delta$ should be included as an 
explicit degree of freedom in the Lagrangian. d) Also in \cite{FeMe}
it is suggested that the data in \cite{Arndt} yields a too large
$\sigma$ term when analyzed with HBChPT.

Further work along these lines is still in progress and
a more detailed presentation with
additional and more complete results will be presented elsewhere.

\begin{figure}[t]
\begin{center}
    \includegraphics[scale=0.6]{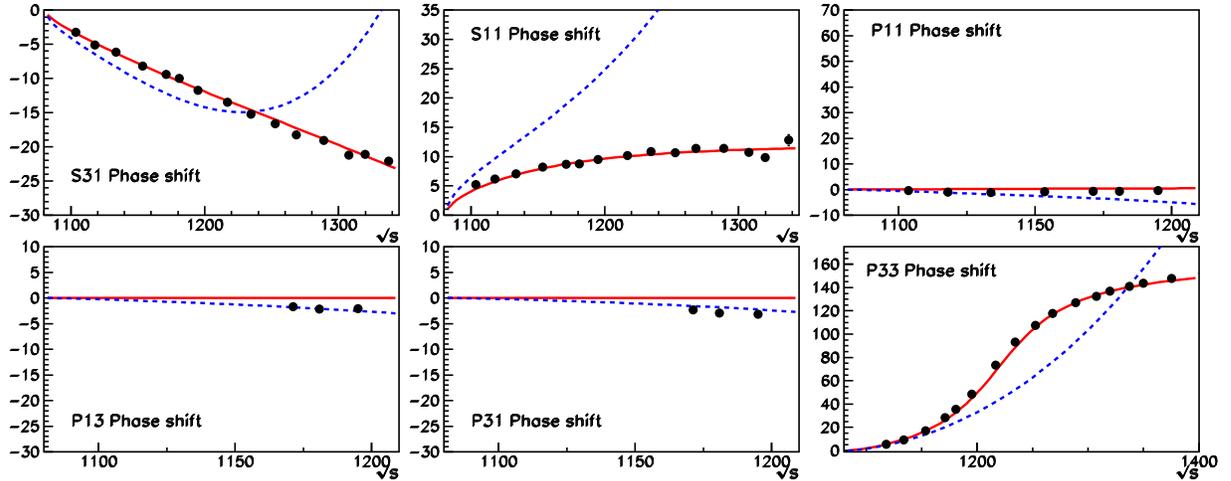}
\end{center}
\vspace{-1.5cm}
\caption{Phase shifts of $\pi$-N scattering. The data comes from \cite{Arndt}.
The continuous line is the IAM fit and the dashed line is the result
of HBChPT  using the chiral parameters of \cite{Mojzis}.}
\label{fig:largenenough}
\end{figure}



\begin{thebibliography}{9}
\footnotesize
\bibitem{HBChPT} E. Jenkins and A. V. Manohar, \PL{B255} (1991) 558;
V. Bernard et al,\NP{B388} (1992) 315. G. Ecker, Czech. J. Phys. {44} (1994) 405.

\bibitem{Gasser} J. Gasser, M. E. Sainio and A. Svarc, \NP{B307} (1988) 779.

\bibitem{Georgi} H. Georgi, \PL{B240} (1990) 447.

\bibitem{Mojzis} M. Mojzis, Eur. Phys J. C 2 (1998) 181.

\bibitem{Meissner} V. Bernard, N. Kaiser and U. -G. Mei{\ss}ner, 
Nucl. Phys. A615 (1997) 483;  N.Fettes, U.-G. Mei{\ss}ner and 
S. Steininger, Nucl. Phys. A (1998) 199.

\bibitem{others} A. Datta and S. Pakvasa, \PR{D56} (1997) 4322; P.J. Ellis 
and H.-B. Tang, \PR{C57} (1998) 3356.

\bibitem{70s} J.L. Basdevant, Fort. der Phys. 20 (1972) 283.
\bibitem{IAM} T. N. Truong, \PRL{661} (1988)  2526;\PRL{67} (1991) 2260;
A. Dobado, M.J.Herrero and T.N. Truong, \PL{B235} (1990) 134 ; 
A. Dobado and J.R. Pel\'aez,\PR{D47} 4883 (1993); \PR{D56} (1997) 3057.
\bibitem{oop} J.A. Oller, E. Oset and J.R. Pel\'aez, Phys. Rev. Lett. 80 (1998)
3452 ; Phys. Rev. D59 (1999) 074001; and  hep-ph/9909556 in this proceedings.
F. Guerrero and 
J. A. Oller, Nucl. Phys. B537 (1999) 459. 
\bibitem{or} E. Oset and A. Ramos, Nucl.Phys. A635 (1998) 99. 

\bibitem{Arndt}  R. Arndt et al. nucl-th/9807087.
SAID online-program.(Virginia Tech Partial-Wave Analysis Facility).
Solution SP99, http://said.phys.vt.edu. 

\bibitem{FeMe}  P. B\"uttiker and  U.-G. Mei{\ss}ner,
hep-ph/9908247.


\end{thebibliography}
\end{document}